\documentclass[manuscript]{aastex}

\slugcomment{Planetary Science Journal}

\shorttitle{138175}
\shortauthors{Jewitt}

\begin{document}


\title{138175 (2000 EE104) and the Source of Interplanetary Field Enhancements}


\author{David Jewitt$^{1,2}$}
\affil{$^1$ Department of Earth, Planetary and Space Sciences,
UCLA, 
595 Charles Young Drive East, 
Los Angeles, CA 90095-1567\\
$^2$ Dept.~of Physics and Astronomy,
UCLA, 
430 Portola Plaza, Box 951547,
Los Angeles, CA 90095-1547\\
}
\email{jewitt@ucla.edu}

\begin{abstract}
We present the first optical observations taken to characterize the near-Earth object 138175 (2000 EE104).  This body is associated with Interplanetary Field Enhancements (IFEs), thought to be caused by interactions between the solar wind magnetic field and solid material trailing in the orbit of the parent body.  Based on optical photometry, the radius (in meters) and mass (in kilograms) of an equal-area sphere are found to be $r_n = 250(0.1/p_R)^{1/2}$ and $M_n = 10^{11}(0.1/p_R)^{3/2}$, respectively, where $p_R$ is the red geometric albedo and density $\rho$ = 1500 kg m$^{-3}$ is assumed.  The measured colors are intermediate between those of C-type (primitive) and S-type (metamorphosed) asteroids but, with correction for the likely effects of phase-reddening, are more consistent with a C-type classification than with S-type. No evidence for co-moving companions larger than $\sim40(0.1/p_R)$ m in radius is found, and no dust particle trail is detected, setting a limit to the trail optical depth $\tau \le 2\times10^{-9}$.  Consideration of the size distribution  produced by impact pulverization  makes it difficult to generate the  mass of nanodust (minimum 10$^5$ kg to 10$^6$ kg) required to account for IFEs, unless the size distribution is unusually steep.  Furthermore, impact pulverization timescales for source objects of the required size are  much longer than the dynamical timescale. While the new optical data do not definitively refute the hypothesis that boulder pulverization is the source of IFEs, neither do they provide any support for it.
\end{abstract}


\section{INTRODUCTION}
Apollo asteroid 138175 (2000 EE104) was discovered on 2000 March 11 in observations taken as part of the Catalina Sky Survey (Christensen et al.~2019).   Its osculating orbital elements (semimajor axis $a$ = 1.0042 AU, eccentricity $e$ = 0.293, inclination $i$ = 5.24\degr), guarantee close dynamical entanglement with the Earth. As well as having a semimajor axis close to Earth's, Connors et al.~(2014) noted that the $q$ = 0.710 AU perihelion of 138175 is close to the $a_V$ = 0.723 AU semimajor axis of  Venus, such that the object can interact strongly with both planets.  Indeed, 138175 passed $\sim$0.01 AU from Earth in 1998 and 1999, and will pass $\sim$0.001 AU from Venus in 2251 (Connors et al.~2014).  Batted between the planets and currently in a near resonance with Earth, 138175 is also classified as a potentially hazardous asteroid.

Observations from space show magnetic field disturbances in the solar wind that are correlated with the orbit of 138175 (Lai et al.~2017).  These disturbances, known as IFEs (Interplanetary Field Enhancements) are characterized by an increase in the field strength $\ge$25\%, a duration $\ge$10 minutes, and the absence of systematic rotation in any of the resolved components of the field (Lai 2014). The latter condition eliminates the passage of flux ropes carried in the solar wind.   First identified in association with the planet-crossing asteroid 2201 Oljato (Russell et al.~1984, Lai et al.~2014), the IFEs have been tentatively interpreted as caused by loading of the magnetic field by nano-scale charged dust particles distributed along the asteroid orbit.  One conceptual model is as follows.  Large boulders are released from the parent asteroids by impact or other processes and spread around the orbit of the parent asteroid, much like the large-particle trails responsible for meteoroid streams. These boulders become secondary sources of fine dust perhaps  through the action of sublimation but more likely from micrometeorite bombardment.  Nanodust released from these secondary sources is charged photoelectrically and then mass-loads the passing  magnetic field, causing a localized pile-up of field lines that is detected by spacecraft as an IFE (Jones et al.~2003b).  The mass of dust needed to significantly load the solar wind field was first  crudely estimated as 10$^8$ to 10$^9$ kg (Lai et al.~2014).   However, these mass estimates are very uncertain and could be high by a factor 10$^3$ (Lai et al.~2017), reducing the dust mass to 10$^5$ kg to 10$^6$ kg.  In the following, we will adopt the smaller mass estimates from Lai et al.~(2017) while keeping in mind the possibility that the dust mass might be much larger.  We note that IFE events are detected at rates about 10 year$^{-1}$ near heliocentric distances $r_H \sim$ 1 AU.

Proposed explanations for the IFE phenomenon remain tentative, but the relation between  IFEs and the orbits of asteroids 2201 Oljato and 138175 (Russell et al.~1984, Lai et al.~2014, 2017) cannot be ignored.  Technically, the proposed link to ejected solids would qualify both Oljato and 138175 as members of the ``active asteroid'' population (Jewitt et al.~2015).  Activity in these objects has been traced to a surprisingly diverse range of physical processes (impact, sublimation, rotational instability, thermal fracture and more), many of which might operate in 138175.  Short-period comet 122P/de Vico has also been reported as a source of IFEs  (Jones et al.~2003a), strengthening the idea that IFEs might be caused by dust loading.  

Our aims in this paper are 1) to draw attention to the IFE phenomenon, 2) to provide the first observational characterization of 138175 and 3) to assess the suggested impact-produced nanodust mechanism.

\section{OBSERVATIONS}
We observed using the Keck 10 m telescope atop Mauna Kea in Hawaii on UT 2018 April 19 and 2019 October 29.  On both occasions we used the LRIS camera (Oke et al. 1995) which permits simultaneous observations in red and blue wavelength channels, separated by a dichroic beamsplitter.  On the blue side, the B filter has central wavelength 4370\AA~and FWHM = 878\AA~while, on the red side, the R filter has central wavelength 6417\AA~and FWHM = 1185\AA. We used the ``460'' dichroic which has 50\% transmission at 4900\AA.  Observations in 2018 were taken at the native pixel scale of 0.135\arcsec~pixel$^{-1}$, while in the 2019 observations we binned the data $2\times2$ to obtain a pixel scale of 0.270\arcsec~pixel$^{-1}$.  The seeing was about 1.0\arcsec~Full Width at Half Maximum (FWHM) on UT 2018 April 19 and 0.8\arcsec~FWHM on UT 2019 October 29.  Observations on the former date were marred by extensive but variable cloud cover and we use observations from this date only to measure the apparent magnitude in the R filter using field stars as a reference.  Observations on the latter date benefited from photometric skies, as established from repeated star measurements in our own data.  On the latter date we obtained photometric calibration from observations of Landolt (1992) stars.  As a precaution, we checked to see if the finite opening time of the large LRIS camera shutters might introduce a significant fractional uncertainty  to the short exposures used on the Landolt star.  We found no non-linearity greater than 1\% in the relevant exposure range, from 2 s to 6 s.

\section{RESULTS}

\subsection{Photometry} 
Observations on UT 2018 April 19 were taken through variable cloud and in 1.0\arcsec~seeing.  The object appeared point-like, save for slight trailing in some images owing to autoguider interruptions by cloud.  From a long series of non-sidereally tracked 400 s integrations, we selected the three least-obscured images and obtained photometry relative to field stars, the latter calibrated from the Sloan DR 14 catalog (Blanton et al.~2017).  

At the time of observation, 138175 was moving with respect to the stars at $\sim$70\arcsec~hour$^{-1}$ in Right Ascension and $\sim$30\arcsec~hour$^{-1}$ in Declination, causing the images of stars to be trailed by 7.5\arcsec.  As a result, small photometry apertures cannot be used because they exclude light from the trailed stars. Conversely, photometry using large apertures suffers from excessive sky noise and background object contamination.  After some experimentation, we found that a 6.8\arcsec~radius photometry aperture, with background subtraction from a contiguous annulus having outer radius 23\arcsec, was sufficient to obtain $\sim$1\% photometry.   The large sky annulus was used to effectively reject contributions from passing background field galaxies.   We used the transformation equations from Sloan system magnitudes to Bessel magnitudes given by Jordi et al.~(2006).  

\begin{figure}
\epsscale{.95}
\plotone{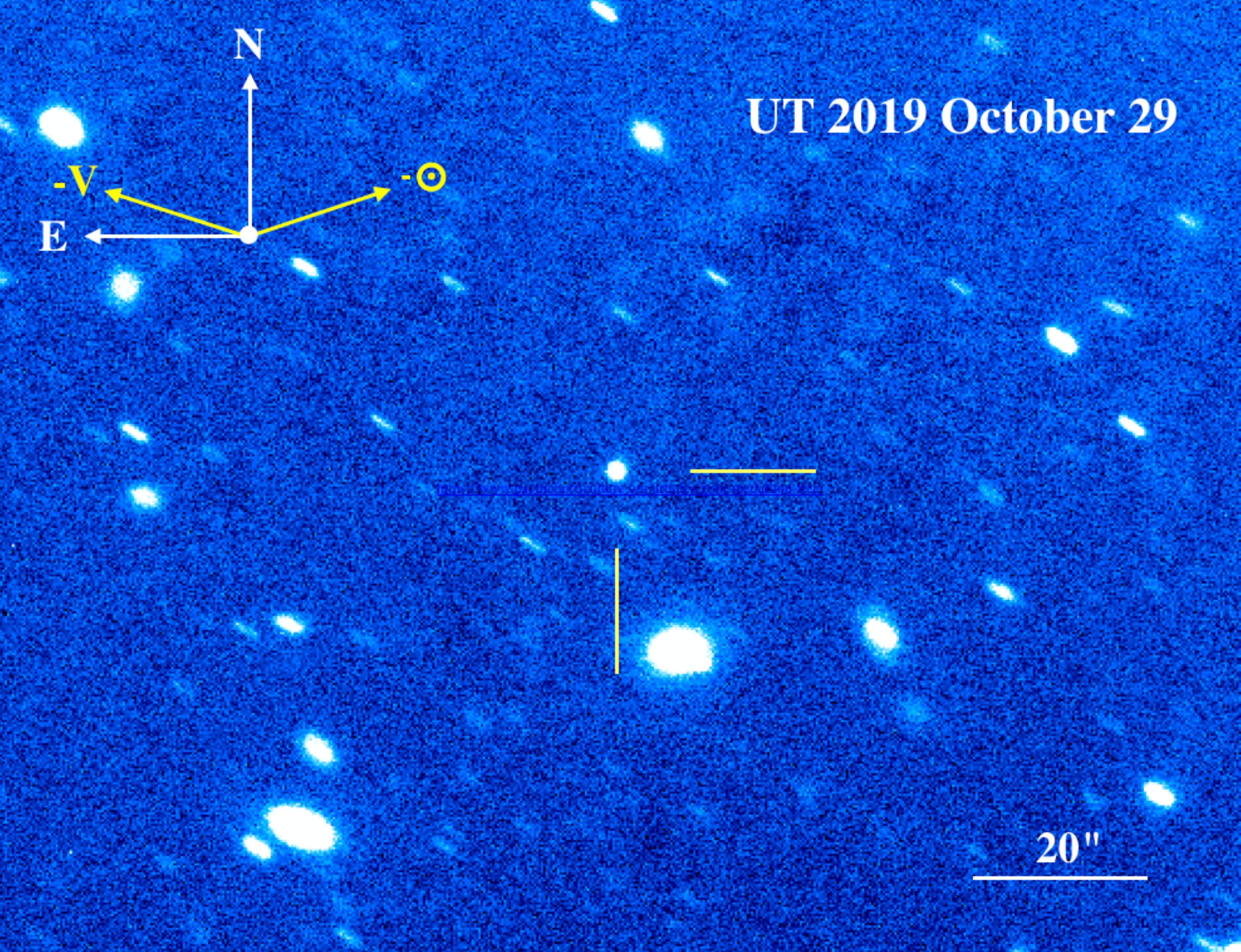}
\caption{Single 280 s, R-filter Keck image from UT 2019 October 29.  North is up, East is left, and $-\odot$ and $-V$ mark the projected anti-solar and negative heliocentric velocity vectors, respectively.   The telescope is tracked non-sidereally to follow the motion of 138175, seen at the center of this image sub-frame.   \label{october29}}
\end{figure}

Observations on UT 2019 October 29 were taken at new Moon (illumination $\sim$3\%), under photometric conditions in 0.8\arcsec~seeing.  The object again appeared point-like, with no trace of coma or trail (Figure \ref{october29}).  We  obtained photometry using a projected circular aperture and sky annulus, as above.  Photometric calibration was obtained using observations of the Landolt (1992) standard stars L101-342 and PG 0918+029A, measured using the same apertures.   Three images of 280 s integration were obtained in  the B and R filters, with results listed in Table (\ref{photometry}).  Since the B and R data were obtained simultaneously in pairs, there is no possibility of contamination of the measured object color by rotational effects.  

The absolute red magnitude, $H_R$,  was computed using

\begin{equation}
H_R = R - 5\log_{10}(r_H \Delta) - f(\alpha)
\label{H}
\end{equation}

\noindent where $R$ is the apparent magnitude, $r_H$ and $\Delta$ are the heliocentric and geocentric distances, respectively, and $f(\alpha)$ is the phase function.   The phase function of 138175 is unmeasured.  We assume a C-type asteroid phase function, expressed in the ``HG'' formalism of Bowell et al.~(1989) by parameter $g$ = 0.15.  The resulting determinations of $H_R$ from the 2018 and 2019 datasets are consistent, within the uncertainties (Table \ref{photometry}).   If we instead adopted an S-type asteroid phase function, then $H_R$ would be 0.22 magnitudes fainter, giving a measure of the importance of the systematic errors on $H_R$.

We next estimate the radius of 138175 from the absolute magnitude, but note that the result is additionally uncertain owing to the unknown value of the geometric albedo, $p_R$.  In the absence of any measurement, we adopt $p_R$ = 0.1, recognizing that primitive asteroids generally have  albedos smaller by a factor of two to three, while metamorphosed asteroids can have albedo higher by a similar factor (Masiero et al.~2011).  The nucleus mass is $M_n = 4/3\pi \rho r_n^3$, where $\rho$ is the bulk density, for which we take the measured average density of asteroids, $\rho$ = 1500 kg m$^{-3}$ (Hanus et al.~2017).  We find

\begin{equation}
r_n = 250 (0.1/p_R)^{1/2}, ~~~M_n = 10^{11} \left(\frac{0.1}{p_R}\right)^{3/2}
\end{equation}

\noindent with $r_n$ in m and $M_n$ in kg.  

\subsection{Color}
\begin{figure}
\epsscale{.7}
\plotone{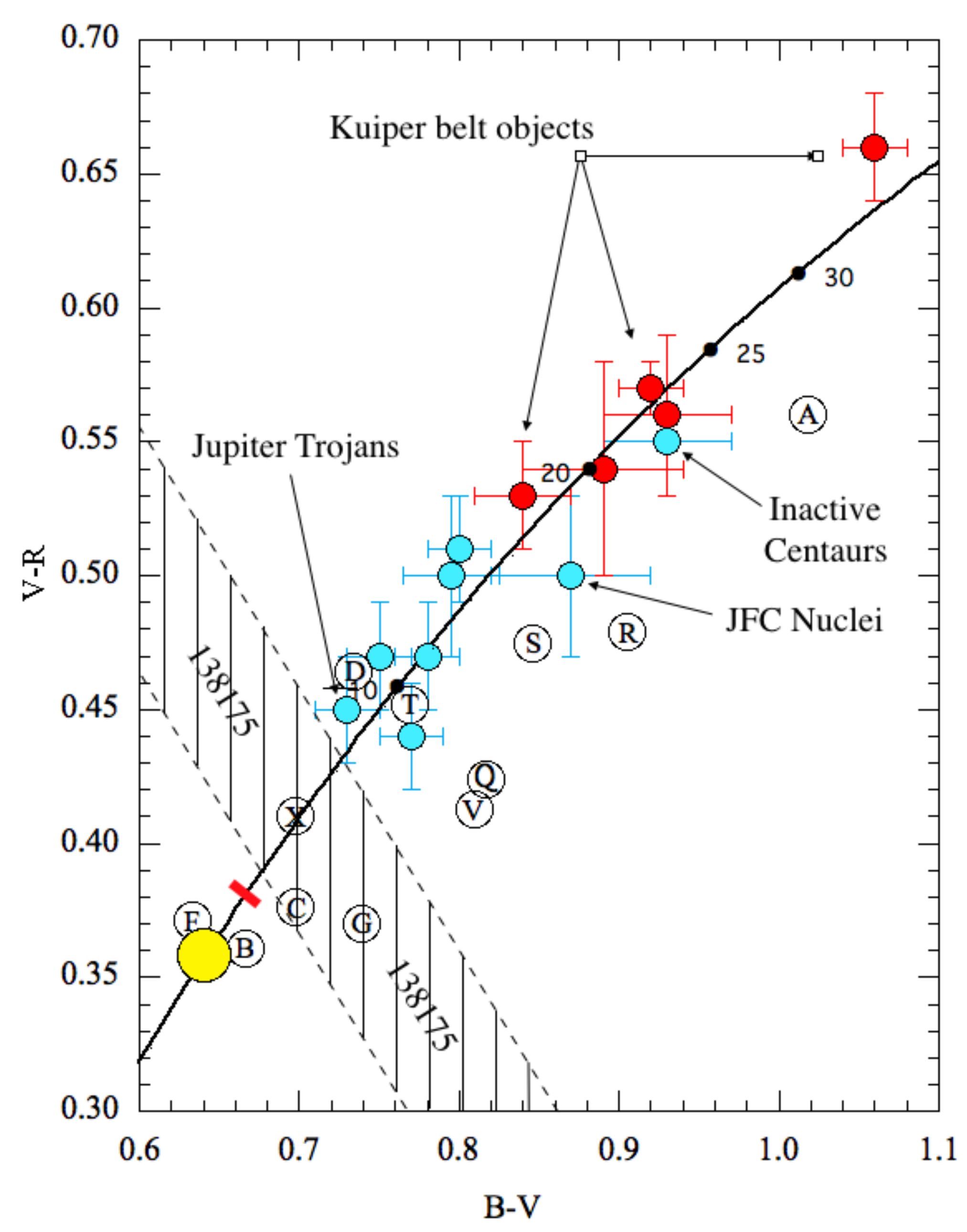}
\caption{Optical color-color plot showing 138175 (diagonal shaded band) in relation to other solar system small-body populations.  The color of the Sun is marked by a large yellow circle.  Red circles are different dynamical sub-types of Kuiper belt object while blue circles are inner and middle-solar system objects (from Jewitt 2015).  Circles containing letters mark the colors of asteroid spectral types from Dandy et al.~(2003).  The diagonal black arc shows the locus of points having fixed spectral reflectivity gradient (\%/1000\AA), marked numerically.  The red bar shows the approximate zero phase angle color, corrected as described in the text. \label{color_color}}
\end{figure}

The measured color, B-R = 1.16$\pm$0.04, is slightly redder than the color of the Sun (B-R = 0.99$\pm$0.02, Holmberg et al.~2006), giving 138175 a reflectivity gradient $S'$ = 8\% (1000 \AA)$^{-1}$.    In the absence of a V-filter measurement we cannot place 138175 on a  B-V vs.~V-R  color-color diagram, but we can identify the  range of allowed locations, shown in the hatched region of Figure (\ref{color_color}), together with the mean colors of other solar system objects from Dandy (2003) and Jewitt (2015).  The limits of the box mark the $\pm$1$\sigma$ uncertainties on the color of 138175.  Evidently, this object is intermediate in B-R between the C-class and S-class asteroids and is, surprisingly, consistent with the mean B-R measured in Jovian Trojan asteroids and D-types, generally.  We are reluctant to over-interpret this observation, however, noting that the color was determined at a much larger phase angle (76.8\degr) than is typical for asteroid color determinations.  The large phase angle potentially renders the measurement of 138175 susceptible to ``phase reddening'', in which the phase function is itself wavelength-dependent because of scattering effects in the regolith.   In the near-Earth population, the largest phase reddening coefficients, $\gamma$, are found in the A ($\gamma$ = 0.49$\pm$0.07 \%/1000\AA/degree), Q ($\gamma$ = 0.07$\pm$0.02 \%/1000\AA/degree) and X ($\gamma$ = 0.07$\pm$0.02 \%/1000\AA/degree) spectral classes (Perna et al.~2018).  While A-class asteroids are comparatively rare, the Q- and X-types together constitute $\sim$25\% to 35\% of the near-Earth objects (Table 2 of Perna et al.~2018);  it would not be unlikely for 138175 to belong to one of these highly phase-reddened classes.     We show the effect of de-reddening the color of 138175  to $\alpha$ = 0\degr~using $\gamma$ = 0.07 \%/1000\AA/degree, as appropriate for the Q and X classes, as a red bar in Figure (\ref{color_color}).  After accounting for phase reddening the figure shows that it is more likely that 138175 is a member of the C-class asteroids than of the S-class.  

\subsection{Limits to Co-Moving Companions} 
 
No co-moving point-source companions are evident in the data, over a square field of view 9\arcmin~on a side ($1.4\times10^5$ km at the distance of 138175).  The detection of such objects is limited by the signal-to-noise ratio (SNR) on clear-sky regions of the field of view, and by overlap with trailed field objects in others.  Slightly diffuse field galaxies are particularly problematic in this regard. The impact of overlap grows with magnitude such that, by R $\sim$ 24, we estimate that the probability of missing a co-moving companion rises to $\sim$50\%.  This is $\sim$ 4 magnitudes fainter than 138175, meaning that bodies having the same albedo as 138175 and six times smaller (i.e.~radius $\sim$ 40 m) have 50\% chance of being missed in our data.  

\subsection{Limits to Diffuse Material} 
The limits to visibility of diffuse near-nucleus material depend on the sky noise, but also on the geometrical distribution of the material.  The latter is a function of  many unknowns, including the particle size and velocity distributions and the angular dependence of their source.  For this reason, it is not possible to establish a single quantitative limit to the presence of near-nucleus material.  However, given the present context, we focus on the likely distribution of large ejected particles that could be the secondary source of nanoscale dust.  

\begin{figure}
\epsscale{.77}
\plotone{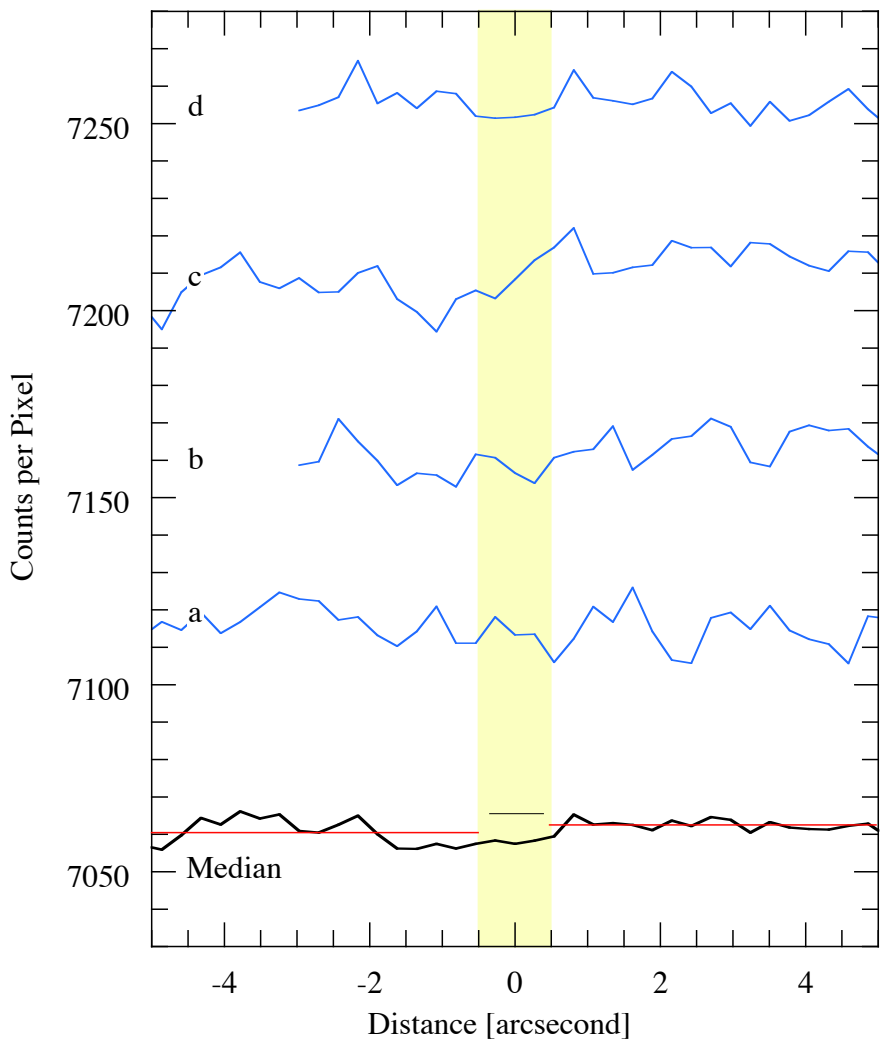}
\caption{Surface brightness measured perpendicular to the projected velocity vector in strips each 8.1\arcsec~wide.  Strips a through d (blue lines) are centered 9.5\arcsec~E,  9.5\arcsec~W, 17.6\arcsec~W~and 25.7\arcsec~W from the nucleus along position angle 74.2\degr.  The plots are vertically offset by multiples of 50 counts pixel$^{-1}$, for clarity. The thick black line shows the median of the other profiles. The shaded yellow region is 1\arcsec~in width and centered on the nucleus.  \label{strips}}
\end{figure}

Large particles   from comets and active asteroids are typically ejected at small  relative velocities  and are confined close to the orbital plane of the parent body, forming thin, quasi-linear ``trails'' in the plane of the sky.  The median trail width in comets is $w \sim$ 27\arcsec (Reach et al.~2007), while dust trails from active asteroids are only $w \lesssim$1\arcsec~wide (Jewitt et al.~2015).  We searched for evidence of such  trails by combining the R filter data from UT 2019 October 29, for a total exposure of 840 s.  Then, we computed surface brightness profiles perpendicular to the projected negative velocity vector (position angle $\theta_{-V}$ in Table \ref{geometry}), averaging over  segments 8.1\arcsec~wide along the $\theta_{-V}$ direction,  in order to improve the SNR.  In practice, we found it difficult to identify large regions free from the effects of trailed background objects.  Figure (\ref{strips}) shows four of the cleanest profiles labeled ``a'' through ``d'', on a linear scale where  7060 data-number units corresponds to  red surface brightness 20.4 magnitude arcsec$^{-2}$.   The four profiles have been vertically offset in the plot for clarity and we have removed pixels (in profiles b and d) in which contamination from field objects was unavoidable.  At the bottom, we show the median of the four plots, ``a'' through ``d''.  The profiles provide no evidence for a dust trail of any width along the projected orbit of 138175.

We computed the mean and standard deviation of the mean signal to the east and west of the nucleus, finding $\Sigma_E = 7060.4\pm0.8$ and $\Sigma_W = 7062.3\pm0.4$.  We take the larger fractional error to estimate a 5$\sigma$ limit to the surface brightness of any narrow dust band of 5$\sigma$ = 28.5 magnitudes arcsec$^{-2}$.   We convert this to line-of-sight optical depth, $\tau$, using (Jewitt et al.~2018)

\begin{equation}
\tau = 1.3\times10^{11} \left(\frac{r_H^2}{p_R \Phi(\alpha)}\right) 10^{0.4\Delta R}
\end{equation}

\noindent where $\Delta R = R_{\odot} - R_1$ is the difference between the magnitude of the Sun ($R_{\odot} =$ -27.06) and the apparent magnitude of one square arcsecond of the dust, which is numerically equal to the surface brightness.  Also, $\Phi(\alpha) = 10^{-0.4 f(\alpha)}$.  Substituting $r_H$ = 1.014 AU (Table \ref{geometry}), $p_R$ = 0.1, $\Phi(76.8\degr) = 0.06$,  and $R_1 \ge$  28.5, we find $\tau \le 2\times10^{-9}$.  For comparison, the measured optical depths of comet dust trails are $10^{-9} \lesssim \tau \lesssim 10^{-8}$ (Ishiguro et al.~2009).

One square arcsecond at $\Delta$ = 0.346 AU (Table \ref{geometry}) corresponds to an area $A = 6\times10^{10}$ m$^2$ and so contains a scattering cross-section $A\tau \le 125$ m$^2$ in dust.  This limits the presence of particles in each square arcsecond to a single object of radius $( A\tau/\pi)^{1/2} =$ 6 m in radius, or less, or to a larger number of smaller particles.  Of course, the data are consistent with the presence of no particles at all.  Lastly, despite the fact that Lai et al.~(2017) reported IFEs over several decades ending in 2012, we cannot formally reject the possibility that the responsible activity had stopped by the time of our observations in 2017 and 2019.  
 
\section{DISCUSSION}
We focus our discussion on the charged nanodust loading hypothesis for the origin of IFEs (Russell et al.~1984, Jones et al.~2003a,b, Lai 2014, Lai et al.~2014, 2017).  In so doing, we note that a surprising number of processes are capable of generating dust and debris from asteroids, including impact,  rotational instability, sublimation of trapped volatiles, thermal fracture, desiccation, and electrostatic lofting (Jewitt et al.~2015).  Indeed,  these varied processes are the main drivers of mass loss in the active asteroids population.  However,  large masses of nanodust are more likely to be produced by impact than by the other mechanisms listed above because of the effects of interparticle cohesion.  Cohesive forces vary with particle size, $a$, in proportion to $a^{-n}$, with $n \sim$ 1 (Sanchez and Scheeres 2014).  Therefore, nanodust experiences cohesive forces that are orders of magnitude larger than typical for the 10 $\mu$m and 100 $\mu$m sized particles commonly observed in the active asteroids.  Whereas large cohesive forces are easily overcome by the huge shock pressures generated by impact, they are less readily exceeded by the other mechanisms, impeding the production of nanodust.  Consequently, we assume that IFE nanodust, if present, must be produced by impact into parent bodies.  

We briefly consider the size of body needed to generate a characteristic mass of nanoparticles,  $M_0$ kg.  If the  body is pulverized, the required object radius is just $r_b = [3 M_0/(4\pi \rho f_{np})]^{1/3}$, where $\rho$ is the object density and $f_{np}$ is the fraction of the target mass pulverized into nanoparticles.  We take the measured average density of asteroids, $\rho$ = 1500 kg m$^{-3}$ (Hanus et al.~2017).  With complete pulverization, $f_{np}$ = 1, and assuming  $M_0 = 10^{5}$ to 10$^6$ kg (Lai et al.~2017), we find  $r_b$ = 2.5 m to 5.5 m.  

However, more realistically, impact destruction of a body results in fragments with a distribution of sizes.  Experiments show that this distribution can be approximated  by a differential power law, in which the number of fragments with size between $a$ and $a+da$ is $n(a)da = \Gamma a^{-q}da$, where $\Gamma$ and $q$ are constants (e.g.~Buhl et al.~2014, Flynn et al.~2020).  Assuming such a distribution, the fraction of the mass contained in nanoparticles, $f_{np} = M_0/M_T$, is

\begin{equation}
f_{np}  = \frac{\int_{a_1}^{a_2} a^{3-q} da}{\int_{a_1}^{a_3} a^{3-q} da}
\label{fraction}
\end{equation}

\noindent in which $M_T$ is the total mass, the nanoparticles extend in radius from $a_1$ to $a_2$, the largest particle in the distribution has radius $a_3$, and the particle density is assumed to be independent of radius.  For illustration, we assume that nanoparticles have radii from $a_1$ = 10$^{-9}$ m to $a_2$ = 10$^{-7}$ m, and we take $a_3$ = 250 m, the radius estimated for 138175 based on optical photometry.  

\begin{figure}
\epsscale{.68}
\plotone{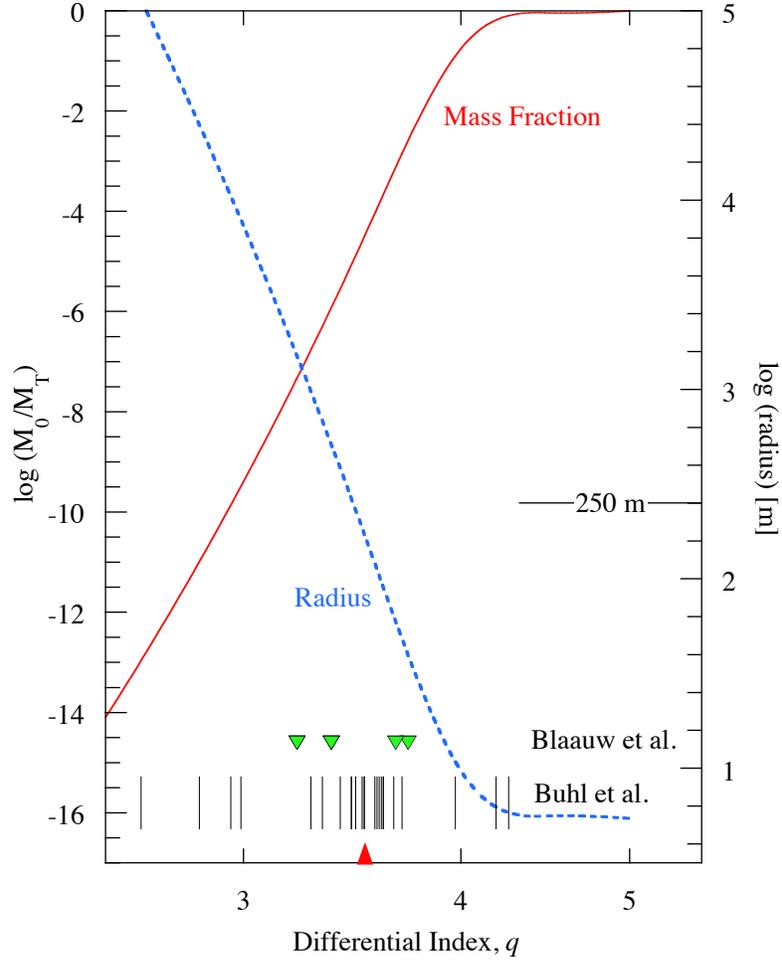}
\caption{Plot showing the fraction of ejected mass contained within nanoparticles (red line, left-hand axis) vs.~the differential power law size index, $q$, from Equation (\ref{fraction}).  The right hand axis (dotted blue line) shows the radius of the progenitor body needed to supply 10$^6$ kg of nanoparticles, the typical mass needed to explain IFEs according to Lai et al.~(2017).  At the bottom, vertical black lines show experimental impact determinations of $q$ by Buhl et al.~(2014) while green-filled triangles show median values of $q$ determined from radar observations of meteor streams by Blaauw et al.~(2011).  The red arrow shows the Dohnanyi (1969) index. The estimated radius of 138175 is marked.   \label{mass_plot}}
\end{figure}

 What is the appropriate value of $q$?  Power law distributions with $q > 4$ place most of the mass in small particles, while those with $q < 4$ hold the mass at the large end of the size distribution.  This is shown in Figure (\ref{mass_plot}), where we plot solutions of Equation (\ref{fraction}) vs.~$q$. The nominal Dohnanyi (1969) model of a collisional cascade has $q$ = 3.5, but the relevance of this to the case of impact-generated debris is not obvious (see also Pan and Schlichting 2012).  More relevant are measurements of the flux of interplanetary nanoparticles (size range $\sim$ 10 nm to 100 nm) near 1 AU, which give $q \sim 3.6$ (Schippers et al.~2014).   Separately, the size distributions in six meteoroid streams  were measured, albeit for much larger particles ($\gtrsim$ 300 $\mu$m), using radar observations by Blaauw et al.~(2011).  Values, for the Geminids ($q$ = 3.22), Quadrantids (3.37), Arietids (3.22), Eta Aquariids (3.73), Orionids (3.37) and South Delta Aquariids (3.67) are shown in Figure (\ref{mass_plot}) as green-filled diamonds.  The median value is $q$ = 3.4 (mean 3.43$\pm$0.09). We also plot  29 experimental impact determinations of $q$ summarized by Buhl et al.~(2014) (two experiments by Cintala et al.~(1985) impacting into water ice have been excluded from consideration, since 138175 at 1 AU is unlikely to preserve internal ice).   The median experimental value is $q$ = 3.5 (mean 3.37$\pm$0.09), again surprisingly close to the Dohnanyi value.  The scatter of the experimental determinations reflects the range of target materials (basalt, granite, Sandstone, Diorite), impact speeds (0.3 km s$^{-1}$ to 7.3 km s$^{-1}$) and specific energies (0.4 J kg$^{-1}$ to 50 J kg$^{-1}$) employed in the laboratory impacts.  All but two of the 29 experiments yield $q < 4$, for which $f_{np}$ = 1 is a poor approximation.   With index $q$ = 3.4 to 3.6, Equation (\ref{fraction}) gives $f_{np} \sim$ 10$^{-5.5}$ to 10$^{-3.8}$.  

To generate $M_0$ = 10$^5$ kg to 10$^6$ kg of nanodust would require the impact destruction of a parent boulder of mass $M_T = M_0 f_{np}^{-1}\sim$ 10$^{8.8}$ kg to  10$^{11.5}$ kg (corresponding to boulder radii  $r_b \sim$ 50 m to 370 m for assumed density $\rho$ = 1500 kg m$^{-3}$).  These  values of mass and radius are alarmingly close to, or even larger than, the mass and radius of 138175 obtained from our photometry ($M_n \sim 10^{11}$ kg, $r_n$ = 250 m).   

The collisional lifetime of boulders in the orbit of 138175 presents another concern.  The  planet-crossing  asteroid population  is comparatively rarefied and the collision rate is dominated by impacts with main-belt asteroids  near aphelion (Bottke et al.~1993).  However, the  aphelion distance of 138175 is only 1.29 AU,  so that collisions with main-belt objects are impossible and the resulting collisional timescales are long;  boulders in the orbit of 138175 and 50 m to 370 m in size have collisional lifetimes  $\sim 10^8$ yr to $\sim 10^9$ yr (Bottke et al.~1993). This  is vastly longer than the $\sim 10^2$ yr timescale for scattering of the orbit   by frequent and strong gravitational interactions with Earth and Venus (Connors et al.~2014) and may even exceed the $\sim 10^8$ yr median dynamical lifetime of planet-crossing asteroids.  Consequently, unless  the planet-crossing projectile population is severely under-estimated, collisional pulverization of boulders released from 138175 must occur too slowly to match the frequency with which IFEs occur.  

We conclude that the production of IFEs cannot reasonably be attributed to impact pulverization of source bodies released from 138175, unless the size distribution of ejecta is substantially  steeper ($q \gtrsim 4.2$, from Figure \ref{mass_plot}) than determined experimentally or in nature, and/or the population of small impactors is much larger than we think.  We cannot a-priori reject these latter possibilities and therefore we cannot definitively reject the pulverized boulder hypothesis.  On the other hand, our optical data  provide no observational support for it.

%
%
%
%

%

\clearpage 

\section{SUMMARY}
We present optical observations of planet-crossing asteroid 138175 (2000 EE104) taken using the Keck 10 m telescope.  This object is associated with repeated  localized  interplanetary magnetic field enhancements  (IFEs, Lai et al.~2017), posited to result from the release of  nanodust  by impact erosion of parent boulders distributed along the orbit.

\begin{enumerate}

\item The R-filter absolute magnitudes measured at two widely-separated epochs are in good agreement,  at $H_R$ = 19.68$\pm$0.03.  The radius (in meters) and mass (in kg) of an equal-area sphere are $r_n = 250 (0.1/p_R)^{1/2}$ and $M_n = 10^{11} (0.1/p_R)^{3/2}$, respectively, where $p_R$ is the red geometric albedo and density $\rho$ = 1500 kg m$^{-3}$ is assumed.  
The optical color measured at phase angle $\alpha$ = 77\degr~is B-R = 1.16$\pm$0.04,  intermediate between the colors of asteroids in the C-type (primitive) and S-type (metamorphosed) spectral classes.  Allowance for likely phase reddening renders 138175  more likely to be C-type than S-type.

\item No evidence for a dust trail was detected, limiting the line-of-sight optical depth to $\tau \le 2\times10^{-9}$.  Neither were discrete sources (boulders) detected, with a limiting magnitude R = 24, corresponding to boulder radius  40 $(0.1/p_R)^{1/2}$ m.

\item Given empirical ejecta size distributions, the boulder source mass needed to generate 10$^5$ kg to 10$^6$ kg of nanodust (the minimum mass estimates according to Lai et al.~2017) is an implausibly large 10$^{8.8}$ to 10$^{11.5}$ kg, rivaling or exceeding the mass of 138175 itself.   Moreover, the collisional lifetime of these large boulders exceeds the dynamical stability timescale. Either the ejecta size distribution is unexpectedly steep (power-law index $q \ge$ 4.2) or the IFEs have another cause.

\end{enumerate}

\acknowledgments

We thank Hairong Lai, Jing Li, Yoonyoung Kim, Pedro Lacerda and the two anonymous reviewers of this paper for their comments.

{\it Facilities:}  \facility{Keck Observatory}.

\clearpage

\begin{deluxetable}{lcccrccccr}
\tablecaption{Observing Geometry 
\label{geometry}}
\tablewidth{0pt}
\tablehead{ \colhead{UT Date and Time}      & \colhead{$\nu$\tablenotemark{a}}  & \colhead{$r_H$\tablenotemark{b}}  & \colhead{$\Delta$\tablenotemark{c}} & \colhead{$\alpha$\tablenotemark{d}}   & \colhead{$\theta_{\odot}$\tablenotemark{e}} &   \colhead{$\theta_{-V}$\tablenotemark{f}}  & \colhead{$\delta_{\oplus}$\tablenotemark{g}}   }
\startdata
2018 Apr 19  15:22 - 16:03         & 222.0 & 1.174 & 0.774 & 57.8 & 98.3 & 280.0 & ~0.4 \\
2019 Oct 29  14:16  - 14:28        & 108.8 &   1.014 &  0.346  & 76.8 & 288.0 &  74.2 & -2.6 \\

\enddata


\tablenotetext{a}{True anomaly, degrees}
\tablenotetext{b}{Heliocentric distance, in AU}
\tablenotetext{c}{Geocentric distance, in AU}
\tablenotetext{d}{Phase angle, in degrees}
\tablenotetext{e}{Position angle of the projected anti-Solar direction, in degrees}
\tablenotetext{f}{Position angle of the projected negative heliocentric velocity vector, in degrees}
\tablenotetext{g}{Angle of Earth above the orbital plane, in degrees}

\end{deluxetable}

\clearpage

\begin{deluxetable}{lcccccccccc}
\tablecaption{Photometry
\label{photometry}}
\tablewidth{0pt}
\tablehead{
 UT Date  &  \colhead{$N$\tablenotemark{a}} &   \colhead{$B$} &  \colhead{$R$}&  \colhead{$H_R$}&  \colhead{$B-R$} }

\startdata
2018 Apr 19   & 3R & --- & 21.55$\pm$0.03 & 19.68$\pm$0.03 & ---  \\
2019 Oct 29 & 3B, 3R  & 21.25$\pm$0.03 & 20.09$\pm$0.03 & 19.69$\pm$0.03 & 1.16$\pm$0.04

\enddata


\tablenotetext{a}{Number of images, by filter}

\end{deluxetable}

\clearpage

\clearpage

\clearpage

\clearpage

\clearpage


%

\end{document}